# Revised JNLPBA Corpus: A Revised Version of Biomedical NER Corpus for Relation Extraction Task

Ming-Siang Huang [1,2],

elephant52381@iis.sinica.edu.tw

Po-Ting Lai [3,4],

s102062802@m102.nthu.edu.tw

Richard Tzong-Han Tsai [5]*

thtsai@csie.ncu.edu.tw

Wen-Lian Hsu [3,4]*

hsu@iis.sinica.edu.tw

[1] Bioinformatics Program, Taiwan International Graduate Program, Institute of Information Science, Academia Sinica, Taipei, Taiwan

[2] Institute of Biomedical Informatics, National Yang Ming University, Taipei, Taiwan

[3] Department of Computer Science, National Tsing-Hua University, Hsinchu, Taiwan

[4] Intelligent Agent Systems Laboratory, Institute of Information Science, Academia Sinica, Taipei, Taiwan

[5] Intelligent Information Service Research Laboratory, Department of Computer Science and Information Engineering, National Central University, Taoyuan, Taiwan

* Corresponding author

# Abstract

The advancement of biomedical named entity recognition (BNER) and biomedical relation extraction (BRE) researches promotes the development of text mining in biological domains. As a cornerstone of BRE, robust BNER system is required to identify the mentioned NEs in plain texts for further relation extraction stage. However, the current BNER corpora, which play important roles in these tasks, paid less attention to achieve the criteria for BRE task. In this study, we present Revised JNLPBA corpus, the revision of JNLPBA corpus, to broaden the applicability of a NER corpus from BNER to BRE task. We preserve the original entity types including protein, DNA, RNA, cell line and cell type while all the abstracts in JNLPBA corpus are manually curated by domain experts again basis on the new annotation guideline focusing on the specific NEs instead of general terms. Simultaneously, several imperfection issues in JNLPBA are pointed out and made up in the new corpus. To compare the adaptability of different NER systems in Revised JNLPBA and JNLPBA corpora, the $F_1$-measure was measured in three open sources NER systems including BANNER, Gimli and NERSuite. In the same circumstance, all the systems perform average 10% better in Revised JNLPBA than in JNLPBA. Moreover, the cross-validation test is carried out which we train the NER systems on JNLPBA/Revised JNLPBA corpora and access the performance in both protein-protein interaction extraction (PPIE) and biomedical event extraction (BEE) corpora to confirm that the newly refined Revised JNLPBA is a competent NER corpus in biomedical relation application. The revised JNLPBA corpus is freely available at iasl-btm.iis.sinica.edu.tw/BNER/Content/Revised_JNLPBA.zip.

# 1. Background and Motivation

The application of natural language processing (NLP) is useful for biomedical scientists to retrieve valuable information from numerous biological publications. Among bio-NLP tasks, biomedical named entity recognition (BNER) is the fundamental but critical one to conquer. Hence, there are many BNER corpora developed for this purpose. For instance, Arizona Disease Corpus (AZDC) [1] is one of the well-known bio-NER corpora. The AZDC corpus contains 793 PubMed abstracts for the disease entity recognition task which is less discussed. Moreover, GENIA corpus [2] is collected by retrieving abstracts associated with specific MEDLINE query terms such as "*human*",

"*blood cells*" and "*transcription factors*". Based on the GENIA ontology [3], the recruited knowledge is well-organized in this corpus. The release of the GENIA corpus promoted text-mining studies in the field of molecular biology and it serves as the seed for several tasks where truthful training and test sets can be constructed.

However, the earlier NER corpora are gradually unsatisfactory to corresponding applications since the development of information extraction studies progresses rapidly ever than before. Take AZDC corpus as an example, the dataset is improved at the entity level and generated the specific identifiers from MeSH or OMIM databases for disease normalization [4]. In contrast to the more detailed corpus revision, combination of trivial concepts to hyper definitions is another direction to promote ideal corpus. The JNLPBA 2004 shared task [5] is derived from five superclasses in the GENIA corpus while the entities are named protein, DNA, RNA, cell line and cell type, respectively.

Due to the complexity of the gene name nomenclature and composition [6], there are several ways to present gene entities in different gene mention corpora. In JNLPBA corpus, the gene mentions are divided into protein, DNA and RNA which included not only specific gene names but also corresponding general sequence information which are seldom referred to as interested entities in biological fields. Another gene mention dataset, GENETAG [7] rules out the general sequence mentions in annotations and considers protein, DNA and RNA as the same entity type since the inclusion allows to reduce the unnecessary disambiguation process. More recently, Gene and Protein Related Object (GPRO) task [8] builds up the GPRO corpus with proposing two types of entity mention definitions: GPRO entity mention type 1 and type 2. All the entities covered in GPRO entity mention type 1 are able to be normalized while entities in type 2 are not. Nevertheless, the gene mentions in GPRO entity mention type 2 represent to the protein family terms or multiple gene mentions rather than the general bio-entities. Despite the standard of gene mentions annotations has been transformed from general gene mentions to more compatible annotation for relation extraction. However, the standard of original corpora still have their reputation, therefore, even recently published works still have to use these corpora to evaluate their approaches (despite the corpus annotations do not fit real application). Another common issue of earlier corpora is that they have either smaller scale or lower annotation consistency because they have to spend more efforts in collecting data and defining annotation standard. Therefore, if a proposed algorithm uses the corpus for evaluation, and get poor results, it is hard to distinguish whether the lower performances come from their algorithm designs or not.

In the paper, we developed a revised version of JNLPBA corpus. We believe that it would be an alternative option for those who use the JNLPBA corpus and get lower performance. Meanwhile, several odd situations found in original JNLPBA are pointed out with relevant cases to emphasize the necessity of corpus revision. It is especially helpful for those who cannot figure out whether the lower performances stand for the inappropriate algorithm designs or not. To measure the efficacy of using Revised JNLPBA, two experiments are conducted. Three widely mentioned NER systems (BANNER [9], Gimli [10] and NERsuite [11]) are introduced in both experiments to evaluate the corpora. In the first experiment, NER systems are trained on both corpora and tested in the corresponding test set. The performance can be a fair reference to judge the annotation consistency of each corpus. To further examine the adaptability of corpus after revision, we integrate five Protein-Protein Interaction Extraction (PPIE) and five Biomedical Event Extraction (BEE) corpora for the validation. All gene mentions in above relation extraction corpora are the prediction targets while the two JNLPBA corpora are utilized for training, respectively.

By the comparisons between the Revised JNLPBA and before, users can select the preferable corpus for their projects according to the needs. In summary, the Revised JNLPBA corpus engages the purities of annotation for researchers who tend to focus on the specific gene mention identification instead of the DNA/protein sequences or the mentions of gene/protein fragments. Moreover, the elusive entity boundaries in the original corpus are also found and improved. Without the interference of the non-specific terms and the boundary confusion, the defects hidden in the desired module is easier to be verified. Finally, the Revised JNLPBA broadens the corpus availability to explore the advanced relation extraction task which the included gene mentions show similar annotation criteria.

## 2. JNLPBA Annotations

In this section, we summarize the imperfections of current JNLPBA corpus and raise the explanations for why we consider that these types of annotations are inappropriate. The homogeneity of JNLPBA corpus is concerned since the entity types from the superclass of GENIA ontology are not excluded the partial components in protein, DNA and RNA. Hence, the contents may present a bunch of noises due to different biological definitions are forced to be clustered together. For example, the partial protein regions such as "*motif*" and "*domain*" are not considered as entities in general biological NER tasks.

Furthermore, given the increasing interest in applying BNER on Protein-Protein Interaction Extraction (PPIE) and Biomedical Event Extraction (BEE) tasks, the annotation standards are not compatible with these tasks. Below, we give some examples of problems found in the JNLPBA corpus. First, many general terms were recruited as entities, such as:

"A construct termed PFP5a DNA containing -795 bp DNA exhibited the highest CAT protein activity, and ..."

The term "*-795bp*" is the position of the DNA sequence (counting from the transcription start site), which is not annotated in the PPIE and BEE corpora, but was referred to as a DNA in the JNLPBA corpus.

Second, some entities include redundant preceding words that do not exist in the original name in the database. For example, *"truncated RARalpha"* is marked as a protein in JNLPBA, but the term *"truncated"* cannot provide any useful identity information to *"RARalpha"* and therefore should be excluded.

In addition, due to the large size of the corpus, there are some missing annotations and incorrect assignment of BNE type in the JNLPBA corpus.

Moreover, there are some inconsistencies that may cause misunderstandings if one attempts to map the entities back to the corresponding ID in standard databases or organize them into clusters. The inconsistencies can be generally classified into five types, including the recruitments of general terms, unnecessary preceding words, entity type confusion, neglected adjacent clues, and missing annotations. The following paragraphs describe examples regarding each type of the confusions that can be found in the GENIA corpus.

## 2.1 Problem of general terms

Specific entities containing unique names can be clearly recognized in the database or certain groups they belong to, while some entities possess only general properties. It is not appropriate for the latter type to be assigned the same label as the former. So, it is better to remove the so-called general terms or assign other tags to them. The following example expresses the appearance of general terms:

"*Substitution mutations in this **consensus sequence** eliminate binding of the **inducible factor**.*"

*MEDLINE: 97138389*

The **bold terms** are labeled as DNA and protein respectively in GENIA corpus but both of them lack the major features to become specific BNEs.

## 2.2 Unnecessary preceding words

The information beyond entities often provides intrinsic properties or external status to make the entities more intact. The intrinsic information is important to assist the assignments of correct identities, so it is suitable to be recruited as part of entities. For example, "*human*" and "*murine*" can serve as the evidence to separate the same "*IL-2*" into different gene IDs. However, extrinsic information sometimes could depict additional properties that are not helpful when distinguishing these entities.

The following instance serves to illustrate this type of words:

"*Expression of **dominant negative MAPKK-1** prevents NFAT induction.*"

*MEDLINE: 96324400*

"*MAPKK-1*" is the core of **bold terms**, while "*dominant negative*" is regarded as a mutant type of protein. The latter biological term carries no useful message if one tries to find the real source of "*MAPKK-1*". Table 1 gives the numbers of general terms and unnecessary preceding words in JNLPBA corpus.

Table 1. Statistics of general terms and unnecessary preceding words in JNLPBA corpus.

| BNE type | Training set | | | Test set | | |
|---|---|---|---|---|---|---|
| | Num. of total NEs | Num. of general terms | Num. of unnecessary preceding words | Num. of total NEs | Num. of general terms | Num. of unnecessary preceding words |
| cell line | 3,830 | 779 | 498 | 500 | 132 | 55 |
| cell type | 6,718 | 1,006 | 1,757 | 1,921 | 174 | 327 |
| DNA | 9,534 | 4,137 | 780 | 1,056 | 306 | 31 |
| protein | 30,269 | 8,145 | 1,781 | 5,067 | 666 | 166 |
| RNA | 951 | 268 | 84 | 118 | 16 | 7 |

## 2.3 Entity type confusion

Some entity types in the GENIA corpus are related to each other, and this association may sometimes cause misclassification of the entity type. By comparing the following two sentences, it is clear that labels are inconsistent.

"... *that the type II IL-1R does not mediate gene activation in **Jurkat cells**.*"

*MEDLINE: 93252936*

"*...galectin-3 was shown to activate interleukin-2 production in **Jurkat T cells**.*"

*MEDLINE: 96208140*

The two similar ***bold entities*** are labeled as cell line and cell type, respectively, in this instance. However, the core term "*Jurkat*" is a powerful attribution to annotate both of them as cell line.

## 2.4 Neglected adjacent clues

In some cases, the entities are concatenated with a certain type of keywords, so it is easy to clarify the real types of entities. But several observations indicate that a few of the keywords are neglected in the GENIA corpus, such as the example below:

"*The 5' sequences up to nucleotide -120 of the human and murine **IL-16** genes ...*"

Without considering the keyword "*genes*", the **bold term** "*IL-16*" was wrongly classified into the protein type.

## 2.5 Missing annotations

This is hard to avoid when the corpus size is huge. GENIA corpus also suffered from this problem. The following sentence displays missing annotations:

"*Three additional smaller regions show homology to the* **ELK-1** *and SAP-1 genes…*"

According to the GENIA ontology, the **bold term** "*ELK-1*" should be referred to as DNA, but it is missing.

In order to obtain more consistent annotations and a reliable source for both entity linking and relation extraction tasks, we had domain experts carefully check and revise the JNLPNA corpus.

# 3. Annotation Guideline

## 3.1 Annotators

For revising JNLPBA, we recruited two curators. Both had a biological background, curation and natural language processing experiences. Annotator 1 is a Ph.D. candidate with biological, chemical and medical background. Annotator 2 is a full-time master research assistant with biological, chemical and computer science background. Annotator 2 also had extensive curation experiences. Each article was annotated independently by the two annotators. Differences were resolved through discussion.

## 3.2 Annotation tool

The curators used the brat annotation tool to manually revise the annotations of the JNLPBA corpus according to the annotation guideline.

## 3.3 General Rules

**Rule 1: Removing General NEs -** In JNLPBA, sometimes nonspecific molecules, like *"upstream regulatory region"*, *"60-kDa protein"* and *"cytokines"*, are annotated as NEs. However, these terms are too general to be linked to any database ID, like

Entrez ID or UniProt ID. Such general terms usually are not labeled as target NEs in the BRE tasks, and thus, are removed.

**Rule 2: Adding Missing NEs -** Due to the considerable size of JNLPBA, sometimes annotations are missed, which are recovered now.

**Rule 3: Adjusting NE Types -** Sometimes cell type such as *"senescent T cells"* is mistaken as a cell line. Thus, all NE types are reconfirmed.

## 3.4 Specific Rules

There is an important principle that must be elucidated first before explaining the detailed criteria of specific rules. The principle is that, when the target NE and surrounding contexts cannot provide any evidence to support the assignment of NE type, this NE is labeled as protein. For instance, the paper titled *"An essential role for NF-kappaB in human CD34 ( + ) bone marrow cell survival."* is insufficient to discriminate the NE type of *"NF-kappaB"*. As a consequence, *"NF-kappaB"* is labeled as protein type since the molecule which is responsible to carry out the biological function is protein.

**Rule 1: Adjective rule -** Only the adjectives preceding an NE that help biologists identify the NE's are included in the NE boundary. Otherwise, they are removed. Moreover, the preserved adjectives often provide the intrinsic properties of described NEs. For example, *"human"* would be included in *"human GM-CSF gene"* because it is related to the NE's species and can help biologists to determine the NE's identifier. In contrast, "abnormal" would be excluded from *"abnormal blast cells"* since the adjective displays an extrinsic property of target NE.

**Rule 2: Verb rule Ving Event adjective verb rule -** If a protein-Ving or protein-Ved phrase is followed by a general term for the protein type, such as protein or transcription factor, protein-Ving/V-ed [protein general term] should be annotated as one single protein mention. If a protein-Ving or protein-Ved phrase is followed by a specific protein mention, then the first protein and the second protein mention should be annotated separately. For example, *"Octamer-binding proteins"* will be labeled as *"Octamer-binding proteins$_{protein}$"*, but "fibrinogen-binding integrins" will be labeled as *"fibrinogen$_{protein}$-binding integrins$_{protein}$"*.

**Rule 3: Preposition rule -** Preposition can be inside an NE only when the preposition is included in the NE's full name. For example, *"Nuclear factor of* *activated T cells (NFAT)"* can be labeled as *"Nuclear factor of activated T cells$_{protein}$ (NFAT$_{protein}$)"*

**Rule 4: Parenthesis rule -** In general, when an abbreviation inside parenthesis follows its full name, it should be labeled separately. For instance, *"tumor necrosis factor (TNF)"* should be labeled as *"tumor necrosis factor$_{protein}$ (TNF$_{protein}$)"*. However, if a specifier number follows the parenthesis, then the whole chunk should be labeled as an NE. For example, *"interleukin (IL) -2"* should be labeled as *"interleukin (IL) -2$_{protein}$"*.

**Rule 5: Conjunction rule -** Conjunctions can appear inside an NE only when they are part of the NE's full name. E.g., *"signal transducers and activators of transcription 5 (STAT5 )"* is labeled as *"signal transducers and activators of transcription 5$_{protein}$ ( STAT5$_{protein}$ )"*. However, if a conjunction is used to connect more than one separate NE, like *"IL-1, 2, and 15"*, it is labeled as *"IL-1$_{protein}$ , 2$_{right\_partial\_protein}$ , and 15$_{right\_partial\_protein}$"*

**Rule 6: Semantic rule -** If the words preceding or following an NE provide additional semantic information that may help to disambiguate the identifier or type of the NE, they will be included. E.g., *"human gene PAX-5"* is labeled as *"human gene PAX-5$_{DNA}$"*.

**Rule 7: Protein rule -** Protein suffixes that describe part of a protein, like "motif" and "domain", rather than a full protein are not included in the protein NE. In addition, the potential protein NEs usually obtain some similar suffixes to represent properties of protein, e.g. protein, receptor, antigen, antibody, enzyme, (transcription) factor and kinase. Thus, the NEs ending with above keywords should be labeled as protein NEs. Moreover, the molecular mass (e.g., "55 kd" in "55 kd TNFR") is a clue to classify the target NE into protein type rather than DNA or RNA.

**Rule 8: DNA rule -** DNA suffixes that describe the function of DNA sequence, like *"enhancer"* and *"promoter"*, are included as part of DNA. More specifically, when there are unique gene names conjugated with the following cis-element evidences, they tend to be labeled as DNA types: (onco)gene, genome, DNA, locus, allele, promoter, enhancer, LTR, response element, probe and plasmid, e.g. *"AP-1 enhancer element"*, *"bcl-2 oncogene"*, *"gene UL49"* and *"FasL promoter"*. Moreover, the NEs with clear chromosome information are also included in annotation, e.g., human chromosome 11p15, 1p36 and 14q11. In some cases, there is not enough evidence to recognize NE type by its own so the surrounding contexts are taken into account for this situation, e.g. *"Pax-5 encodes the transcription factor BSAP which plays an essential role ..."* In the above sentence, *"Pax-5"* is only a gene symbol name without any clue to judge its NE type. But the verb *"encodes"* provides the evidence to classify *"Pax-5"* since DNA is the only one among the five NE types to encode the permanent sequence of a functional protein. Lastly, there is a special expression profile in biological literature to describe the cloned DNA plasmid. In general, the target NE is tight-conjugated with a lowercase *"p"* in the head of NE, e.g. *"pCD41"*, *"pIL-5 cDNA"*.

**Rule 9: Cell rule -** The target NEs which match the following features are referred to as cell_line: 1. NEs described with obvious cell line symbol, e.g. *"Hela"*, *"Hep2"* and *"A549"*; 2. general cell name, cellular morphological or cellular functional description ending with *"cell line"* or *"clone"*, e.g., *"T cell line"*, *"granulocytic clones"* and *"monocytic cell line"*.

The NEs below are annotated as cell_type: the names are mentioned with specific cell type, cellular morphology or cellular functional ending with *"cell"*, *"progenitor"* or *"precursors"*, e.g., *"thymocytes"*, *"hematopoietic cells"* and *"myeloid precursors"*.

**Rule 10: Complex rule -** If a complex is expressed as *"<Protein>/<Protein>"*, it will be treated as one protein. E.g., *"TCR/CD3"* is labeled *"TCR/CD3$_{protein}$"*.

**Rule 11: Amino or DNA sequence rule -** Amino acids and DNA sequences are not labeled as NEs. For example, *"WGATAR consensus motifs"* and *"GGAAAGTCCC"* are not included in the corresponding NE lists.

**Rule 12: Group/family protein -** A protein family is a group of proteins which have similar functions, and a protein complex is a high-level structure consisting of more than one protein. Although, protein families and complexes do not appear in Entrez or UniProt databases, these entities are still very important for relation extraction. So, we include them in Revised JNLPBA.

## 3.5 Inter-annotator Agreement (IAA) Analysis

To evaluate the consistency of the annotation, we used Cohen's kappa coefficient. As shown below, $\kappa$ is the kappa value. $P_0$ is the relative observed agreement among annotators, and $P_e$ is the hypothetical probability of chance agreement.

$$\kappa = \frac{P_0 - P_e}{1 - P_e}$$

There are two curators participating in the two-stage annotations. After the first stage, the two curators would discuss annotation disagreements and then start the second stage annotation. The kappa values are 79.5% and 91.4% in the first and the second stage, respectively, which suggests a high level of agreement.

## 4. Experiment Results

We design two experiments to evaluate the effects of using Revised JNLPBA corpus. In the first experiment, we compare the performances of commonly-used NER systems trained on JNLPBA and Revised JNLPBA respectively. In the second experiment, we train the NER systems on the two editions of JNLPBA and evaluate their performances on Protein-Protein Interaction Extraction (PPIE) and Biomedical Event Extraction (BEE) corpora. PPIE datasets include LLL [12], AImed [13], BioInfer [14], IEPA [15] and HPRD50 [16]. BEE datasets include BioNLP 2013 ST GRO, GE, GRN, CG and PC datasets [17]. We removed all non-gene and cell-related NE annotations from these datasets and combined them into the BRE corpus.

### 4.1 Evaluation Metrics

The performance is given in terms of $F_1$-measure and is calculated by using the evaluation script from JNLPBA.

### 4.2 BNER Systems

Three BNER systems were used for comparison, including BANNER [9], Gimli [10], and NERsuite [11]. We selected these systems because they are available BNER systems and achieved state-of-the-art performances on either JNLPBA or GENETAG. All of them are based on Conditional Random Fields (CRF). The following table summarized the characteristics of these systems.

Table 2. The configurations of the NER systems.

|  | **BANNER [9]** | **Gimli [10]** | **NERsuite [11]** |
|---|---|---|---|
| **Model/Toolkit** | CRF/MALLET | CRF /MALLET | CRF/CRFsuite |
| **Label set** | BIO | BIO | IOBES |
| **Tokenization** | Simple rule | GDep | GENIATagger |
| **Features** | | | |
| Word | Y | Y | Y |
| Chunk | Dragon toolkit | GDep | GENIATagger |
| Lexicon | Y | Y | - |
| Morphological | Y | Y | Y |
| Orthographic | Y | Y | Y |
| POS | Dragon toolkit | GDep | GENIATagger |
| Stem | Dragon toolkit | GDep | GENIATagger |
| **Others** | | | |
| Abbreviation | Y | Y | - |
| Ensemble | - | Y | - |
| Parentheses | Y | Y | - |

## 4.3 Experiment 1

In this experiment, JNLPBA training and test set were used for compared systems. Table 3 shows the performances of different approaches on the test set. We suspect that lower score of BANNER might be that the feature selection in the BANNER system was based on the BioCreative II GM dataset rather than the JNLPBA dataset.

The inconsistencies in the JNLPBA, likely due to annotators with different annotation criteria, create a bottleneck on the BNER performances. To alleviate any negative effects bringing by this problem, we revised the dataset basis on the annotation guideline. The corresponding performances are shown in Table 3. Generally, the overall performances of NER systems can reach at least 10% higher in Revised JNLPBA than the original one.

Table 3. The performances of the NER systems on JNLPBA and Revised JNLPBA.

| System | JNLPBA | | | Revised JNLPBA | | |
| --- | --- | --- | --- | --- | --- | --- |
| | Precision | Recall | F-score | Precision | Recall | F-score |
| BANNER | 66.65 | 69.34 | 67.97 | 89.11 | 73.51 | 80.56 |
| Gimli | 72.85 | 71.62 | 72.23 | 91.33 | 82.84 | 86.88 |
| NERSuite | 69.95 | 72.41 | 71.16 | 89.13 | 83.41 | 86.17 |

## 4.4 Experiment 2

In this experiment, the two JNLPBA corpora were used for training NER open-source systems. Subsequently, PPIE and BEE corpora are served as test set to evaluate the performances. The F-scores of the systems were shown in Table 4 and 5. The lower performance of BNER systems stems from the fact that there is no general consensus regarding PPIE and BEE annotations. The definitions of BNE boundaries differ among PPIE, BEE, and the Revised JNLPBA/JNLPBA corpora. For example, in JNLPBA and Revised JNLPBA, "human Myt1 kinase" is designated as a BNE. However, in the BioInfer corpus, only "Myt1" is included in the BNE.

According to the performance of PPIE based on different training corpora, LLL and IEPA get similar but lower effects no matter which training corpus adapted. The results are explainable because the abstracts in LLL corpus are mainly collected from bacterial domains while the chemical is the major issue of IEPA. In contrast, the performance of AImed, Bioinfer, and HPRD50 are relatively higher with both training resources and the NER systems trained in Revised JNLPBA corpus perform averagely better than their corresponding counterparts which trained in original one.

In the BEE results, NER systems only reach around 30% performance with two training corpora editions in GRO and GRN. By inspecting the data composition of the two corpora, GRO included general protein terms and functional protein fragments as entities while GRN is derived from BioNLP-ST 2011 BI and LLL corpora which present bacteria relevant article as the dominant contents. The distinct biological domain and the inclusion of general entities frustrated the system performances which are trained on the Revised JNLPBA. On the other hand, the better performances can elucidate the entity annotation styles are adaptable for NER systems trained on the Revised JNLPBA when GE, CG and PC corpora serve as the test set in comparison. In average, NER systems trained on the revised edition can reach 5% higher compared with the original ones.

**Table 4.** The performances (in %) of the NER systems on the PPIE corpora.

| | **NER Systems Trained on JNLPBA** | | | | | | | | | | | | | | |
|---|---|---|---|---|---|---|---|---|---|---|---|---|---|---|---|
| | **LLL** | | | **AImed** | | | **BioInfer** | | | **IEPA** | | | **HPRD50** | | |
| | NS | GIM | BNR | NS | GIM | BNR | NS | GIM | BNR | NS | GIM | BNR | NS | GIM | BNR |
| P | .369 | .500 | .354 | .526 | .547 | .511 | .649 | .678 | .634 | .365 | .468 | .366 | .517 | .534 | .503 |
| R | .263 | .22 | .297 | .648 | .61 | .643 | .487 | .472 | .513 | .283 | .314 | .289 | .638 | .593 | .585 |
| F | .307 | .306 | .323 | .581 | .577 | .569 | .556 | .556 | .567 | .319 | .375 | .323 | .571 | .562 | .541 |
| | **NER Systems Trained on Revised JNLPBA** | | | | | | | | | | | | | | |
| | **LLL** | | | **AImed** | | | **BioInfer** | | | **IEPA** | | | **HPRD50** | | |
| | NS | GIM | BNR | NS | GIM | BNR | NS | GIM | BNR | NS | GIM | BNR | NS | GIM | BNR |
| P | .372 | .585 | .634 | .68 | .697 | .687 | .779 | .816 | .801 | .433 | .514 | .459 | .609 | .602 | .634 |
| R | .148 | .161 | .271 | .658 | .631 | .657 | .522 | .510 | .551 | .323 | .355 | .332 | .623 | .621 | .583 |
| F | .212 | .253 | .380 | .669 | .663 | .672 | .625 | .628 | .653 | .370 | .420 | .385 | .616 | .611 | .607 |

**Table 5.** The performances (in %) of the NER systems on the BEE corpora.

| NER Systems Trained on JNLPBA | | | | | | | | | | | | | | | |
|---|---|---|---|---|---|---|---|---|---|---|---|---|---|---|---|
| | GRO | | | GE | | | GRN | | | CG | | | PC | | |
| | NS | GIM | BNR | NS | GIM | BNR | NS | GIM | BNR | NS | GIM | BNR | NS | GIM | BNR |
| P | .2 | .1993 | .194 | .4011 | .4107 | .4091 | .4589 | .4341 | .4155 | .6489 | .6444 | .6241 | .5447 | .5447 | .5445 |
| R | .3948 | .3742 | .402 | .5896 | .5286 | .6101 | .2926 | .2445 | .2576 | .4984 | .4539 | .4955 | .5874 | .5415 | .6111 |
| F | .2655 | .2601 | .2617 | .4774 | .4622 | **.4898** | .3573 | .3128 | .3181 | .5638 | .5327 | .5524 | .5652 | .5431 | **.5759** |
| **NER Systems Trained on Revised JNLPBA** | | | | | | | | | | | | | | | |
| | GRO | | | GE | | | GRN | | | CG | | | PC | | |
| | NS | GIM | BNR | NS | GIM | BNR | NS | GIM | BNR | NS | GIM | BNR | NS | GIM | BNR |
| P | .2394 | .2395 | .2361 | .484 | .5046 | .5159 | .4231 | .4684 | .4878 | .7329 | .7654 | .7352 | .6637 | .6324 | .6819 |
| R | .3751 | .3617 | .3268 | .5908 | .5673 | .5689 | .1441 | .1616 | .1747 | .5048 | .5138 | .4365 | .6122 | .6065 | .5868 |
| F | .2923 | .2882 | .2741 | .5321 | .5341 | .5411 | .215 | .2403 | .2572 | .5978 | .6148 | .5478 | .637 | .6191 | .6308 |

# 5. Conclusion

The progression of biomedical text mining is urgently required in the era of information explosion. Each mature machine learning model is built on not only the well-designed algorithms but also the reliable validation mechanism. Inconsistent annotations make the learning systems confusion. And It is hard to discriminate the truth hidden in the biological texts since there are various nomenclature forms depending on the distinct domains. With a confidential corpus, researchers can concentrate on improving their own systems rather than clarifying the causes of unexpected performance. In this work, we propose a revised edition of JNLPBA corpus. Several imperfections found in original JNLPBA corpus are pointed out and corrected as much as possible. According to the evaluation of different NER systems, we believe that the corpus performs higher consistency than before after revision process. The further application adaptability of revised JNLPBA is also examined via the open tests on PPIE and BEE corpora. In overall, the revised JNLPBA is competent for the systems which required gene mention entities training for relation extraction purpose. We envision the revised JNLPBA corpus can become another option for the researchers who engage in the BNER or BRE issues.

# References

[1] M. C. Leaman R, Gonzalez G, "Enabling recognition of diseases in biomedical text with machine learning: corpus and benchmark.," in

*Proceedings of the 2009 Symposium on Languages in Biology and Medicine.*, 2009, pp. 82-89.

[2] J. D. Kim, T. Ohta, Y. Tateisi, and J. Tsujii, "GENIA corpus--semantically annotated corpus for bio-textmining," *Bioinformatics,* vol. 19 Suppl 1, pp. i180-2, 2003.

[3] T. O. Jin-Dong Kim, Yuka Teteisi, Jun' ichi Tsujii, "GENIA Ontology," Tsujiilab, University of Tokyo, Tokyo TR-NLP-UT-2006-2, 15 November 2006.

[4] R. I. Dogan, R. Leaman, and Z. Lu, "NCBI disease corpus: a resource for disease name recognition and concept normalization," *J Biomed Inform,* vol. 47, pp. 1-10, Feb 2014.

[5] J.-D. Kim, T. Ohta, Y. Tsuruoka, Y. Tateisi, and N. Collier, "Introduction to the bio-entity recognition task at JNLPBA," presented at the Proceedings of the International Joint Workshop on Natural Language Processing in Biomedicine and its Applications, Geneva, Switzerland, 2004.

[6] H. M. Wain, E. A. Bruford, R. C. Lovering, M. J. Lush, M. W. Wright, and S. Povey, "Guidelines for human gene nomenclature," *Genomics,* vol. 79, pp. 464-70, Apr 2002.

[7] L. Tanabe, N. Xie, L. H. Thom, W. Matten, and W. J. Wilbur, "GENETAG: a tagged corpus for gene/protein named entity recognition," *BMC Bioinformatics,* vol. 6 Suppl 1, p. S3, 2005.

[8] R. O. Krallinger M., Lourenço A., "Evaluation, corpora and analysis of chemical and gene/protein name recognition in patents: the CHEMDNER patents text mining task at BioCreative V. ," *Database,* 2016.

[9] R. Leaman and G. Gonzalez, "BANNER: an executable survey of advances in biomedical named entity recognition," *Pac Symp Biocomput,* pp. 652-63, 2008.

[10] D. Campos, S. Matos, and J. L. Oliveira, "Gimli: open source and high-performance biomedical name recognition," *BMC Bioinformatics,* vol. 14, p. 54, Feb 15 2013.

[11] D. o. I. S. Tsujii Laboratory, The University of Tokyo, "NERsuite - A Named Entity Recognition toolkit," ed, 2012.

[12] C. Nédellec, "Learning Language in Logic - Genic Interaction Extraction Challenge.," in *Proceedings of the Learning Language in Logic 2005 Workshop at the International Conference on Machine Learning.*, 2005.

[13] R. Bunescu, R. Ge, R. J. Kate, E. M. Marcotte, R. J. Mooney, A. K. Ramani, and Y. W. Wong, "Comparative experiments on learning information extractors for proteins and their interactions," *Artif Intell Med,* vol. 33, pp. 139-55, Feb 2005.

[14] S. Pyysalo, F. Ginter, J. Heimonen, J. Bjorne, J. Boberg, J. Jarvinen, and T. Salakoski, "BioInfer: a corpus for information extraction in the biomedical domain," *BMC Bioinformatics,* vol. 8, p. 50, Feb 9 2007.

[15] J. B. Ding, D.; Nettleton, D.; Wurtele, Eve., "Mining MEDLINE: abstracts, sentences, or phrases?," in *Pacific Symposium on Biocomputing*, 2002, pp. 326-37.

[16] K. Fundel, R. Kuffner, and R. Zimmer, "RelEx--relation extraction using dependency parse trees," *Bioinformatics,* vol. 23, pp. 365-71, Feb 1 2007.

[17] C. Nédellec, R. Bossy, J.-D. Kim, J.-J. Kim, T. Ohta, S. Pyysalo, and P. Zweigenbaum, *Overview of BioNLP Shared Task 2013*: Association for Computational Linguistics, 2013.